\documentstyle[12pt]{article}
\newcommand{\DP}{D\bar{\psi}D\psi~}

\newcommand{\be}{\begin{equation}} \newcommand{\ee}{\end{equation}}
\newcommand{\bea}{\begin{eqnarray}} \newcommand{\eea}{\end{eqnarray}}
\newcommand{\bp}{{\bf p}} 
\begin{document} 
\title{
Vacuum Properties of a Non-Local Thirring-Like  Model} \author{D.\ G.\
Barci\thanks{barci@symbcomp.uerj.br} \\ {\normalsize \it Universidade do
Estado do Rio de Janeiro,}\\ {\normalsize \it Instituto de F\'\i sica,
Departamento de F\'\i sica Te\'orica} \\ {\normalsize \it R.\ S\~ao
Francisco Xavier, 524,}\\ {\normalsize \it Maracan\~a, cep 20550, Rio de
Janeiro, Brazil.}       \and
C.M.Na\'on\thanks{naon@venus.fisica.unlp.edu.ar)}\\ {\normalsize \it
Depto. de F\'\i sica. Universidad Nacional de La Plata.}\\ {\normalsize
\it CC 67, 1900 La Plata, Argentina.}\\ {\normalsize \it Consejo
Nacional de Investigaciones Cient\'\i ficas y T\'ecnicas, Argentina.}}

\date{May 1997} \maketitle 
\begin{abstract} 
We use path-integral methods to
analyze the vacuum properties of a recently proposed extension of the
Thirring model in which the interaction between fermionic currents is
non-local. We calculate the exact ground state wave functional of the
model for any bilocal potential, and also study its long-distance
behavior. 
We show that the ground state wave functional has a general factored Jastrow 
form. We also find that it posess an interesting symmetry involving
the interchange of 
density-density and current-current interactions.

\end{abstract}
\pagenumbering{arabic}

\vspace{5cm}

\section{Introduction}

In the last years there has been much interest in the study of
low-dimensional field theories. One of the main reasons for this revival
can be found in striking achievements of the material sciences that have
allowed to build ultranarrow semiconductor structures \cite{PT} in which
the motion of the electrons is confined to one dimension \cite{Voit}.
One important tool for the theoretical understanding of the
one-dimensional (1d) electron system is the Tomonaga-Luttinger (TL)
model \cite{To} \cite{L} \cite{ML}, which can be considered as the
paradigm of Luttinger Liquid (LL) behavior \cite{H} \cite{Voit}. This
model describes a non-relativistic gas of massless particles (the
electrons) with linear free dispersion relation and two-body,
forward-scattering interactions. In a recent work \cite{NRT}, a
non-local Thirring model (NLT) with fermionic currents coupled by
general (symmetric) bilocal potentials was presented, which contains the
TL system as a special case. In that work, the complete non-local
bosonization of the model was presented, calculating in particular the
dispersion relations of the relevant bosonic modes involved in the
system. 

In quantum field theory, the wave functional is not the most usual
quantity to calculate. The reason is that, for the evaluation of
scattering data, the relevant quantities are Green functions.
Nevertheless, with the application of quantum field techniques to model
condensed matter systems, it becomes useful to get the information
contained in the ground state of the system. 

In this paper we address our attention to the vacuum properties of the
above mentioned model. As it is well-known, ground-state wave
functionals (GSWF's) have in general very complex structures. Due to
this fact, their universal behavior has been seldom explored in the
past. Fortunately, in a recent series of papers, an alternative way to
compute GSWF's was presented \cite{LF} \cite{F} \cite{FMS}. By
conveniently combining the operational and functional approaches to
quantum field theories, these authors provided a systematic
path-integral method that, at least in the context of $1+1$ systems,
seems to be more practical than the previously known semiclassical
\cite{semi} and Bethe ansatz \cite{stora} \cite{suther} techniques. We
take advantage of these advances and apply them to shed some light on
the vacuum structure of the NLT. This is a relevant issue for several
reasons. On the one hand, our work can be viewed as an extension of the
path-integral approach to wave functionals to the case in which
non-local interactions are taken into account. On the other hand our
studies clarify the physical content of a model that is interesting by
its own right, due to its direct connection with many-body systems.

The paper is organized as follows. In Section 2 we define the NLT model
and recall the steps that enable to obtain its path-integral
bosonization. In Section 3 we describe the density representation of
wave functionals and combine the results of \cite{NRT} and \cite{FMS} in
order to evaluate the GSWF for the NLT. We get a closed formula that
gives the probability of the vacuum state as a functional, not only of
the density configuration but also of the potentials that bind the
original fermionic particles of the system. This result allows us to
find a non-trivial symmetry of this vacuum with respect to the
interchange of potentials. We also discuss the general electromagnetic
response of the model. In Section 4 we analyze the long-distance
behavior of the GSWF for the NLT. Exploiting the generality of this
model we specialize the results of the previous Section to some
particular potentials. This permits us to make contact with the TL
\cite{To} \cite{L} \cite{ML} and Sutherland \cite{suther} models. In
Section 5 we sketch the Grassmann representation of wave functionals and
show how to implement it in the present context. Although the
mathematical structure for the GSWF is more involved in this
representation, once again we obtain a closed expression for the
probability of the vacuum as a functional of both Grassmann sources and
potentials. Finally, in Section 6 we summarize our main results and
conclusions.

\section{The Model and the Non-Local Bosonization Method}
\setcounter{equation}{0}

We start defining our model by writing the partition function
\begin{equation} 
Z = \int D\bar{\Psi}~D\Psi ~e^{-S}, \label{a}
\end{equation} 
where the action $S$ can be split as 
\begin{equation} 
S =
S_0 + S_{int}, \end{equation} with \begin{equation} S_0 = \int d^2x~
\bar{\Psi} i \raise.15ex\hbox {$/$}\kern-.57em\hbox{$\partial$} \Psi
\label{free} \end{equation} and \begin{equation} S_{int} =
-\frac{g^2}{2} \int d^2x d^2y ~ [V_{(0)}(x,y) J_0(x) J_0(y) +
V_{(1)}(x,y) J_1(x) J_1 (y)] \label{int} 
\end{equation} 
where the
electron field $\Psi$ is written as \[ \Psi = \left( \begin{array}{c}
\Psi_{1} \\ \Psi_{2} \\ \end{array} \right), \] with $\Psi_1$ ($\Psi_2$)
describing right (left) movers.

Concerning the electronic kinetic energy, we have set the Fermi velocity
equal to 1. The interaction piece of the action has been written in
terms of currents $J_{\mu}$ defined as 
\begin{eqnarray} 
J_{\mu} &=&
\bar{\Psi} \gamma_{\mu} \Psi, \nonumber\\ 
\label{,} 
\end{eqnarray}
$V_{(\mu)}(x,y)$ are symmetric bilocal arbitrary potentials describing
the electron-electron (e-e) interactions. The model is not relativistic
except for the special case $V_{(0)}=V_{(1)}$. 

Our first goal is to express the functional integral (\ref{a}) in terms 
of fermionic determinants. To this end we define new currents \\
\begin{equation}
K_{\mu}(x) = \int d^2y~ V_{(\mu)}(x,y)J_{\mu}(y).
\label{KK}
\end{equation}
(Note that no sum
over repeated indices is implied when a subindex $(\mu)$ is involved).
 
The usual procedure in order to match the quartic interaction
between fermions consists in introducing auxiliary fields $A_{\mu}$
so that one can write
\begin{eqnarray}
Z & = &\int D\bar{\Psi}~D\Psi~e^{-S_0 }\nonumber\\
&\int & DA_{\mu}~ \delta [A_{\mu} - K_{\mu}] exp [
\int d^2x~J_{\mu}A_{\mu}]\nonumber\\
\end{eqnarray}
On the other hand, we represent the $\delta$ functionals as integrals of 
exponentials over new fields $C_{\mu}(x)$, thus 
obtaining
\begin{eqnarray}
Z & = & \int  D\bar{\Psi}~D\Psi~DA_{\mu}~e^{-S_0 
}\nonumber\\ 
& &exp {\int d^2x~J_{\mu}A_{\mu}}
\int DC_{\mu} exp[-\int d^2x~(A_{\mu} - K_{\mu})C_{\mu}] \nonumber\\
\label{k}
\end{eqnarray}
At this point one sees that the fermionic piece of the action
(the free part and the terms involving the currents J and K) can be cast
in a local form by defining the ``potential transformed''  fields
\begin{equation}
\bar{C}_{\mu}(x) = \int d^2y~ V_{(\mu)}(x,y) C_{\mu}(y),
\end{equation}
We then get
\begin{eqnarray}
Z& = &\int D\bar{\Psi}~D\Psi~DA~D\bar{C}\nonumber\\
& &exp\{-\int d^2x~[\bar{\Psi}( i \raise.15ex\hbox{$/$}\kern-.57em\hbox
{$\partial$} + (\not\!\! A + \not\!\! \bar{C}))\Psi + \nonumber\\
&+& C_{\mu}(x)A_{\mu}(x)]\}.
\end{eqnarray}
This equation, in turn, suggests the following change of variables
\[
A_{\mu} + \bar{C}_{\mu} = \tilde{A}_{\mu},
\]
\begin{equation}
A_{\mu} - \bar{C}_{\mu} = \tilde{C}_{\mu},
\label{13'}
\end{equation}
giving
\begin{eqnarray}
Z = &\int & D\tilde{A}~D\tilde{C}~det(i\raise.15ex\hbox{$/$}
\kern-.57em\hbox{$\partial$} + \not\!\! \tilde{A})~\nonumber\\
& & exp\int d^2x~d^2y~\{- \frac{b_{(\mu)}(x,y)}{4}
[\tilde{A}_{\mu}(x)-\tilde{C}_{\mu}(x)]
\tilde{C}_{\mu}(y)\},
\end{eqnarray}
where we have defined the inverse potentials $b$ through
the identities
\begin{equation}
\int V_{(\mu)}(x,y) b_{(\mu)}(x,z) d^2x~ = \delta^{(2)}(y-z),
\label{aa}
\end{equation}
In the above expression for $Z$ one sees that, by virtue of the change
of variables (\ref{13'}), the fields $\tilde{C}$ play no direct role in
the fermionic determinant. They are actually artefacts of our method,
whereas the fields $A$ describe the physically relevant bosonic degrees
of freedom. Therefore, the next step is to perform the integrals in
$\tilde{C}$. This can be easily done, as usual, by conveniently shifting
the fields. When this is done one finds a field that describes negative
metric states. In order to agree with Klaiber's operational prescription
we absorb the decoupled ghost partition function in the overall
normalization constant \cite{NRT}. Taking these considerations into
account, and setting from now on $\tilde{A}=A$, one finally gets
\begin{equation} 
Z =  \int DA~ e^{-S'[A]} det ( 
i\raise.15ex\hbox{$/$}\kern-.57em\hbox{$\partial$} +
\not \!\! A ) .
\label{6}
\end{equation}
with
\begin{equation}
S'[A]  =  \int d^2x ~ d^2y\frac{1}{2}
[A_{\mu}(x) b_{(\mu)}(x,y) A_{\mu}(y)]. 
\label{7}
\end{equation}

Thus we have been able to express the partition function for the NLT in
terms of a fermionic determinant. This is a necessary condition to apply
the path-integral approach to non-local bosonization which, combined
with the methods of \cite{LF} and \cite{FMS}, will enable us tu derive
the GSWF for the NLT. An interesting point of our approach is that one
can go further quite a long way without specifying the potentials. This
will be shown in the next Sections, where we shall undertake the
evaluation of the GSWF in the density and Grassmann representations.

\newpage

\section{Ground State Wave Functional in the Density Representation} 

Generally, the wave function can be labeled by the eigenvalues of the
particle number operator. In the case of an N-particle system,
$\psi(x_1,\ldots,x_n)=<\psi|x_1,\ldots,x_n>$, where $|x_1,\ldots,x_n>$
is an eingenstate of the particle density operator $\hat\rho(x)=\hat
c^\dagger(x)\hat c(x)$ with eigenvalue
$\rho(x)=\sum_{i=1}^{N}\delta(x-x_i)$. This representation is called the
density representation and we can label it by $\psi(\rho)=<\psi|\rho>$. 

In the case of a dense system, $\rho(x)$ is a general distribution (not
necessarily $\delta$'s) and $\psi(\rho)$ is a functional of a density
rather than a function.

In a fermionic system, antisymmetrization of the wave function is
supposed. In spite of the fact that $\rho(x)$ is an even distribution in
a dense system, we will see that the Pauli exclusion principle is still
satisfied. 
 
This section is devoted to the calculation of the GSWF
in the density representation, for the non-local
Thirring model, described in the previous Section.

The GSWF is related with the equal-time density correlation function
$<\rho(p)\rho(-p)>$. In references \cite{LF} and \cite{FMS}, it was
shown an interesting relation between the GSWF in the density
representation and the generating functional $Z(Q)$ (see Appendix): 
\be
|\psi_0[\rho]|^2=\int DQ_0 e^{-i\int d{\bf x} Q_0({\bf x}) \rho({\bf
x})} \lim_{Q_0(x)\to Q_0({\bf x})\delta(x_0)} Z(Q_0, Q_1=0) \label{gs}
\ee

So, the first step towards the calculation of $|\psi_0[\rho]|^2$ is to
evaluate $Z(Q_\mu)$. To this aim, we take advantage of equation
(\ref{6}), where the generating functional was written in terms of a
fermionic determinant. Coupling the system minimally to an external
gauge potential $Q_\mu(x)$, we obtain from (\ref{6}): 

\bea
\lefteqn {Z(Q_\mu)=\int  DA ~ 
 Det\left\{ i \not \! \partial+ g \not\! A\right\} \times  }  \\     
       &\times&\exp\left\{ - \frac{1}{2}
\int d^2xd^2y~(A_\mu(x)+\frac{1}{g}Q_\mu(x)) b_{(\mu)}(x-y)
(A_\mu(y)+\frac{1}{g}Q_\mu(y)) \right\} \nonumber
\eea

To calculate the  fermionic determinant, we make a chiral change of
variables in the fermionic fields 

\bea
\psi(x)&=&e^{-g[\gamma_5 \phi(x)+i\eta(x)]} \chi\nonumber \\
\bar{\psi}(x)&=&\bar{\chi} e^{-g[\gamma_5 \phi(x)-i\eta(x)]}
\label{chiral}
\eea 
With this change of variables the measure transforms as
\be
\DP = J_F[\phi,\eta] D\bar{\chi}D\chi
\ee
It is also possible (in 1+1 dimensions),  to split the gauge field in a 
longitudinal plus a transversal component in the following way
\be
A_\mu(x)=\epsilon_{\mu\nu}\partial_\nu\phi+\partial_\mu\eta
\label{longi}
\ee
These changes lead to
\be
Det\left\{ i \not \! \partial+ g \not\! A\right\} =
J_F[\phi,\eta]Det\left\{ i \not \! \partial\right\}
\ee
As it is well-known (see for instance \cite{GS}) the Jacobian associated
to this change of variables is: 
\be
\log{J_F[\phi,\eta] }=\frac{g^2+\alpha}{2\pi}\int d^2x~ \phi\Box\phi
\ee
where $\alpha$ is an arbitrary parameter that can be fixed with gauge 
invariance arguments. 

Putting all this together, we finally find a bozonized generating 
functional 
\be
Z(Q_\mu)=\int D\phi D\eta e^{-S_{eff}(\phi, \eta, Q_\mu)}
\ee
where
\bea
S_{eff}&=&\frac{g^2+\alpha}{2\pi} \int d^2x~ (\partial_\mu\phi)^2
\nonumber \\
&+&\int d^2xd^2y    \left\{ 
b_{(0)}(x-y)\partial_1\phi(x)\partial_1\phi(y)+
b_{(1)}(x-y)\partial_0\phi(x)\partial_0\phi(y) \right\}\nonumber\\
&+&\int d^2xd^2y~ \left\{
b_{(0)}(x-y)\partial_0\eta(x)\partial_1\phi(y)-
b_{(1)}(x-y)\partial_1\eta(x)\partial_0\phi(y) \right\}\nonumber \\
&-&\frac{1}{g}\int d^2xd^2y~ 
\phi(x)(\epsilon_{\mu\nu}\partial_\nu b_{(\mu)}(x-y))Q_\mu(y)\nonumber \\
&-& \frac{1}{g}\int d^2xd^2y~ 
\eta(x)(\partial_\mu b_{(\mu)}(x-y))Q_\mu(y)\nonumber \\
&+&\frac{1}{2g^2}\int d^2xd^2y~ Q_\mu(x)  b_{(\mu)}(x-y))Q_\mu(y)
\label{Sb}
\eea
 
It is simpler to evaluate the generating functional in momentum space. 
To do this, we Fourier transform eq.\ (\ref{Sb}) obtaining: 
\bea
S_{eff} &=&
\int \frac{d^2p}{(2\pi)^2}~[\tilde\phi(p)\tilde\phi(-p)A(p)+
\tilde\eta(p)\tilde\eta(-p)B(p)
+\tilde\phi(p)\tilde\eta(-p)C(p)]+ \nonumber \\
&+&\frac{i}{g}\int \frac{d^2p}{(2\pi)^2}~ \left\{
\tilde\phi(p)(\epsilon_{\mu\nu}p_\nu \hat b_{(\mu)}(p)
\tilde Q_\mu(-p))+
\tilde\eta(p)(p_\mu \hat b_{(\mu)}(p))\tilde Q_\mu(-p))\right\}
 \nonumber \\
&+& \frac{1}{2g^2}\int \frac{d^2p}{(2\pi)^2} \tilde Q_\mu(p))
\hat b_{(\mu)}(p)\tilde
Q_\mu(-p)
\label{seff}
\eea
where
\bea
A(p) &=& \frac{g^2 + \alpha}{2\pi}~ p^2 +
     \frac{1}{2}[\hat{b}_{(0)}(p) p_1^2 +
      \hat{b}_{(1)}(p) p_0^2], \\
B(p) &=& \frac{1}{2}[\hat{b}_{(0)}(p) p_0^2 +
	   \hat{b}_{(1)}(p) p_1^2],\\
C(p) &=& [\hat{b}_{(0)}(p) - \hat{b}_{(1)}(p)] p_0 p_1. 
\eea

To integrate (\ref{seff}), we first decouple the fields $\tilde\phi$ 
and
$\tilde\eta$ by means of 
\bea
\tilde\phi&=&\hat\xi-\frac{C}{2A}\hat\zeta \label{xi} \\
\tilde\eta&=&\hat\zeta \label{zeta}
\eea
and then,  we integrate the quadratic integrals in $\hat\xi$ and 
$\hat\zeta$. 
We thus obtain:
\be
Z(Q_\mu)=\exp\left\{-\int \frac{d^2p}{(2\pi)^2}~ \tilde Q_\mu(p) 
\pi_{\mu\nu}(p)
\tilde Q_\nu(-p)\right\}
\label{gf}
\ee 
where
\be
\pi_{\mu\nu}=
\frac{1}{2\pi}
\frac{p_0^2+\bp^2}{  \{ \frac{g^2 }{\pi }v_0(p)+1  \} {\bf p}^2+
		    \{ \frac{g^2 }{\pi }v_1(p)+1  \} p_0^2 }
\left(
\delta_{\mu\nu}-\frac{p_\mu p_\nu}{p^2}
\right)
\label{pimunu}
\ee
with $v(p)={\cal F}(V(x-x'))$. 

Note that,  for the generating functional $Z(Q_\mu)$ being gauge invariant, 
the polarization tensor $\pi_{\mu\nu}$ must be transversal 
($p_\mu\pi_{\mu\nu}=0$). 
This property  is automatically satisfied for any potential due to the 
tensor  structure of (\ref{pimunu}). 

For the present model being  well defined,  
the potentials must satisfy the following (sufficient) condition:
\be
\left( \frac{g^2 }{\pi }v_0(p)+1  \right)\left( \frac{g^2 }{\pi }v_1(p)+1  
\right) >0
\label{condition}
\ee  

If this relation is not satisfied, the {\em euclidean} $\pi_{\mu\nu}$
may have a pole in the real $p_0$ axis. This corresponds to the
propagation of a runaway mode, breaking the unitarity of the model.
Roughly speaking, relation (\ref{condition}) means that the
density-density interaction and the current-current interaction must be
both repulsive or both attractive (note that it is a sufficient
condition, not necessary).
As a by product we have obtained {\em the exact electromagnetic response
of the system for any potential $v_0$ or $v_1$}. For example, if we
apply an arbitrary electric field to the fermionic system, it will
induce an electric current and a charge density given by (in Minkovsky
space):
\bea
\lefteqn{\rho(p)\equiv<\bar\psi\gamma^0\psi>\label{density}} 
\\&=&-\frac{i}{8\pi^3}
\frac{\bp}{  \{ \frac{g^2}{\pi } v_0(p)+1  \} {\bf p}^2-
		    \{ \frac{g^2 }{\pi }v_1(p)+1  \} p_0^2 } 
\times {\bf E}(p_0,\bp) \nonumber 
\eea
\bea
\lefteqn{
J(p)\equiv<\bar\psi\gamma^1\psi>\label{current}} \\
&=&\frac{i}{8\pi^3}
\frac{p_0}{  \{ \frac{g^2 }{\pi }v_0(p)+1  \} {\bf p}^2-
		    \{ \frac{g^2 }{\pi }v_1(p)+1  \} p_0^2 } 
\times {\bf E}(p_0,\bp) \nonumber 
\eea 
where ${\bf E}=i({\bf p}Q_0-p_0Q_1)$ is the applied electric field.

Equation (\ref{current}) implies that for instantaneous potentials,
the system allows the propagation of free waves with dispersion 
relation 
\be
\omega_0=\pm\sqrt{\frac{ \frac{g^2 }{\pi }v_0(\bp)+1}
{\frac{g^2 }{\pi }v_1(\bp)+1}  } |\bp|
\label{dr}
\ee 
These waves are related with the propagation of the bosonic modes $\hat
\xi$ and $\hat \zeta$ (see eq. (\ref{xi}) and (\ref{zeta})) discussed in
ref.\ \cite{NRT}. We shall return to this issue in the next section, when
we consider the Tomonaga-Luttinger model. In particular, note that
in the special case $v_0(\bp)=v_1(\bp)$, the dispersion relation is
the free one $\omega_0=\pm |\bp|$ ($v_f=1$). 

After this digression, let us now face the evaluation of the GSWF. 
Fourier transforming  equation (\ref{gf}) and using  (\ref{gs})
we have (in euclidean space): 

\be
|\psi_0[\rho]|^2=\int D\tilde Q_0 
e^{-\frac{1}{2\pi}\int d\bp \tilde Q_0(\bp) \rho(\bp)}
e^{-\frac{1}{(2\pi)^2}\int d\bp~ \tilde Q_0(\bp) \bar\Pi_{00}(\bp)
\tilde Q_0(-\bp)} 
\label{wf}
\ee
where
\be
\bar\pi_{00}(\bp)=\int dp_0~ \pi_{00}(p_0,\bp)
\label{pi}
\ee

It is a simple task to integrate equation (\ref{wf}) obtaining
\be
|\psi_0[\rho]|^2=e^{\frac{1}{4}\int d\bp \rho(\bp)
\left(\bar\pi_{00}\right)^{-1} \rho(-\bp)}
\label{result1}
\ee
with $\bar\pi_{00}$ given by (\ref{pi}). 

This result express the probability (not the amplitude) of a 
particular density distribution to be realized in the ground
state of the non-local Thirring model.

We can go further if we suppose that the potentials are 
local in time (as in any non-relativistic model). In this 
case, $v_0$ and $v_1$ are $p_0$-independent and we can integrate 
(\ref{pi}) explicitely. So for any instantaneous potential we have
\bea
\bar\pi_{00}(\bp)&=&\int_{-\infty}^{+\infty} \frac{dp_0}{2\pi}~ 
\frac{\bp}{  (\frac{g^2 }{\pi  }v_0(\bp)+1)  {\bp}^2+
	     (\frac{g^2 }{\pi  }v_1(\bp)+1)  p_0^2}         \nonumber\\
&=& \frac{\pi}{2}\sqrt{\frac{1}{ (\frac{g^2 }{\pi  }v_0(\bp)+1) 
	     (\frac{g^2 }{\pi  }v_1(\bp)+1)}}
|\bp|
\label{piins}
\eea
In this way, from (\ref{piins}) and (\ref{result1}) we obtain 
\be
|\psi_0[\rho]|^2=e^{\frac{1}{2}
\int d\bp 
\rho(\bp)
\sqrt{ (\frac{g^2 }{\pi  }v_0(\bp)+1) 
	     (\frac{g^2 }{\pi  }v_1(\bp)+1)}
\left(\frac{1}{|\bp|}\right)
\rho(-\bp)}
\label{result2}
\ee
This equation is the main result of this section, and gives
the {\em exact
ground state wave functional for the non-local Thirring model } defined
by eqs. (\ref{free}) and (\ref{int}). 

Let us point out some general features of this wave functional. 
 From (\ref{result2}) we see that the vacuum of the theory has the 
non-trivial  symmetry
\be
V_{(0)}(x-y)\longleftrightarrow V_{(1)}(x-y) \label{simetryx}
\ee
This symmetry tells us that the density-density interaction is 
completely 
equivalent to the current-current interaction, provided we are studying
only vacuum properties.  It is clear that it should be broken  by the
excited states of the spectrum since the action (\ref{int}) is not
symmetric.

Another property of (\ref{result2}), is that $|\psi[\rho]|^2$ in the N-
particle  subspace, has a general factored Jastrow form:
\be
|\psi(x_1,\ldots,x_n)|^2=\prod_{i,j}\left[ \varphi(|x_i-x_j|)\right]^\lambda 
\label{jastrow}
\ee
To see this more  clearly, let us rewrite (\ref{result2}) in configuration 
space,
\be
|\psi_0[\rho]|^2=e^{\pi
\int dx_1~dx_2
\rho(x_1)
f(|x_1-x_2|)
\rho(x_2)}
\label{result3}
\ee
with 
\be
f(|x_1-x_2|)=\int \frac{dp}{2\pi}~\sqrt{ (\frac{g^2}{\pi  } v_0(\bp)+1) 
	     (\frac{g^2 }{\pi  }v_1(\bp)+1)} 
\left(\frac{1}{|\bp|}\right)
e^{i \bp(x_1-x_2)}
\label{f}
\ee
We now consider the Fock subspace with fixed $n$ particles and
$n$ holes (antiparticles), since we are working with $<\rho>=0$, 
i.\ e.\ ,  without chemical  potential.  In this circumstance one has
\be
\rho(x)=\sum_{i=1}^{n}\left( \delta(x-x_i)-\delta(x-y_i)\right)
\label{rho}
\ee
where $\{x_i\}$($\{y_i\}$) is  the position of the particles (holes). 

Replacing (\ref{rho}) in (\ref{result3}) we have

\be
|\psi(x_1, \ldots, x_n, y_1, \ldots, y_n)|^2=
\prod_{i<j} e^{2\pi f(|x_i-x_j|)} e^{2\pi f(|y_i-y_j|)}
\prod_{i,j} e^{-2\pi f(|x_i-y_j|)}
\label{result4}
\ee 
that is the general factored form for our model. Clearly, the complexity
of this expression depends on the form of the potentials. In the next
section we will study universal properties of the long-distance
behavior. But, in order to gain confidence and to somehow test our
result, let us analyze the simpler example of a local Thirring model, i.e. 
$v_0=v_1=1$. We then get 
\be
f(|x|)=(\frac{g^2}{\pi}+1){\cal F}
(|\bp|^{-1})=\frac{1}{\pi}(\frac{g^2}{\pi}+1)\ln(|x|)   
\ee
Inserting this expression in (\ref{result4}) we obtain
\be
|\psi(x_1, \ldots, x_n, y_1, \ldots, y_n)|^2=
\frac{ \prod_{i<j}|x_i-x_j|^{2\mu}|y_i-y_j|^{2\mu}}
{\prod_{i,j}|x_i-y_j|^{2\mu}}
\ee
with $\mu=g^2/\pi+1$, which is the correct result \cite{FMS}.

\section{Long distance behavior and the connection with the
Tomonaga-Luttinger and Sutherland models}

In this section we analyze the general long-distance behaviour of the
GSWF. We also specialize the non-local Thirring
model for particular potentials, showing that it contains the
Tomonaga-Luttinger \cite{To} \cite{ML} and the Sutherland \cite{suther}
models as particular cases. 

We have already shown, that in the limit $V_0=V_1=\delta^2(x-y)$, our
model reproduces the local Thirring model. Let us now consider two
non-local potentials of the form: 
\bea V_0&\propto &
|x-y|^\alpha\delta(x_0-y_0) \label{alpha} \\ V_1&\propto &
|x-y|^\beta\delta(x_0-y_0) \label{beta} 
\eea Their fourier transforms
are \cite{gelfand}

\bea
v_0(\bp)&\propto&-2 \sin(\frac{\alpha}{2}\pi)
\Gamma(\alpha+1) |\bp|^{-\alpha- 1}\\
v_1(\bp)&\propto&-2 \sin(\frac{\beta}{2}\pi)\Gamma(\beta+1) 
|\bp|^{-\beta- 1}
\label{v}
\eea

The long distance behavior of $|\psi(\rho)|^2$, is dominated by
$\lim_{x\rightarrow\infty} f(x)$, or equivalenty $\lim_{\bp\rightarrow
0} \tilde f(\bp)$(see eqs.\ (\ref{result3}) and (\ref{f})). This
behavior depends on the different values of $\alpha$ and $\beta$. 

For large $x$ (small $\bp$) we have the following situation, 
\be
\begin{array}{llll}
\left.          \begin{array}{l}
\alpha>-1 \\
 \beta>-1  
		\end{array}
\right\} \rightarrow          
&  \tilde f(\bp)\propto |\bp|^{-
\frac{\alpha+\beta+2}{2}}  &\rightarrow    
f(x)\propto -\mbox{cte.}
|x|^{\frac{\alpha+\beta}{2}}\\ 
\left.          \begin{array}{l}
\alpha>-1 \\
 \beta<-1  
		\end{array}
\right\} \rightarrow           &  
\tilde f(\bp)\propto |\bp|^{-\frac{\alpha+3}{2}} &\rightarrow  
f(x)\propto -\mbox{cte.}
|x|^{\frac{\alpha+1}{2}}\\ 
\left.          \begin{array}{l}
\alpha<-1 \\
 \beta>-1  
		\end{array}
\right\} \rightarrow          &  
\tilde f(\bp)\propto |\bp|^{-
\frac{\beta+3}{2}} &\rightarrow  f(x)\propto -\mbox{cte.}
|x|^{\frac{\beta+1}{2}}\\ 
\left.          \begin{array}{l}
\alpha<-1 \\
 \beta<-1  
		\end{array}
\right\} \rightarrow           &  
\tilde f(\bp)\propto \frac{1}{|\bp|} &\rightarrow 
f(x)\propto ~\mbox{cte.} \ln|x| 
\end{array}
\label{table}
\ee

We note that the Jastrow form (\ref{jastrow}) is kept for all potentials
at long distances, but the physics may be very different depending on
the values of $\alpha$ anb $\beta$. From the first line of
(\ref{table}), we see that for $\alpha>-1$ and $\beta>-1$ we have two
types of long distance limits. If $\alpha+\beta<0$, the wave functional
tends asymptotically to a constant, a typical behavior of an
incompressible fluid. But, if $\alpha+\beta>0$, the wave functional goes
to zero exponentially as $\exp(-\mbox{cte.}
|x-x'|^{\frac{\alpha+\beta}{2}})$, characterizing a confining phase.
Lines two and three of (\ref{table}) also characterize a confining
phase, because this exponential decay of the wave function implies that
asymptotic fermionic states cannot exist. Last but not least, the case
$\alpha<-1,~\beta<- 1$, represent another phase, that can be identified
as the Thirring phase, since the general form of the wave function is
the same as in the local Thirring model with the appropriate
redefinition of the coupling constant (the exponent in the Jastrow wave
function). 
 
Another way of classifying these phases, is through the transport
properties of the systems. That is to say, each phase is associated to a
free propagating mode with different dispersion relations in each phase.
From (\ref{dr}) and (\ref{v}), we can write the long distance dispersion
relations for the free propagation modes as: \be \begin{array}{ll}
\left. \begin{array}{l} \alpha>-1 \\ \beta>-1 \end{array} \right\}
\rightarrow & w_0=\pm |\bp|^{\frac{\beta-\alpha+2}{2}} \\ \left.
\begin{array}{l} \alpha>-1 \\ \beta<-1 \end{array} \right\} \rightarrow
& w_0=\pm |\bp|^{\frac{1-\alpha}{2}} \\ \left. \begin{array}{l}
\alpha<-1 \\ \beta>-1 \end{array} \right\} \rightarrow & w_0=\pm
|\bp|^{\frac{\beta+3}{2}} \\ \left. \begin{array}{l} \alpha<-1 \\
\beta<-1 \end{array} \right\} \rightarrow & w_0=\pm |\bp| \end{array}
\label{drtable} \ee 

Note that these relations distinguish between $V_0$ and $V_1$ ($\alpha$
and $\beta$). This is so because, in order to propagate such modes, not
only the ground state is necessary, but the excited states are necessary
also.

This completes our analysis of the long distance behavior in our
non-local Thirring model. Let us now show, by specifying the potentials,
that this general theory contains some other models already discussed in
the literature related to 1-d strong correlated systems. 

We shall apply the approach developed in previous Sections to the
Tomonaga-Luttinger model \cite{To} \cite{L} \cite{ML}. This model
describes a non-relativistic gas of spinless and massless particles
(electrons) in which the dispersion relation is taken to be linear. The
free-particle Hamiltonian is given by \be H_0 = v_{F} \int dx
\Psi^{\dagger}(x) (\sigma_{3} p - p_{F}) \Psi(x) \label{48} \ee
\noindent where $v_{F}$ and $p_{F}$ are the Fermi velocity and momentum
respectively ($v_{F}p_{F}$ is a convenient origin for the energy scale).
$\sigma_{3}$ is a Pauli matrix and $\Psi$ is a column bispinor with
components $\Psi_1$ and $\Psi_2$ ($\Psi^{\dagger} =
(\Psi_1^{\dagger}~~\Psi_2^{\dagger})$). The function $\Psi_1(x)
~[\Psi_2(x)]$ is associated with the motion of particles in the positive
[negative] $x$ direction. The interaction piece of the Hamiltonian, when
only forward scattering is considered, is

\be H_{int} = \int dx \int dy \sum_{a,b} \Psi^{\dagger}_a (x) \Psi_a(x)
V_{ab}(x,y) \Psi^{\dagger}_b(y) \Psi_{b}(y) 
\label{49} 
\ee

\noindent where $a,b=1,2$, and the interaction matrix is parametrized
in the form

\be
V_{ab} = \left( \begin{array}{cc}
	    v_1 & v_2\\
	    v_2 & v_1
	    \end{array} \right).
\label{50}
\ee

\noindent Using the imaginary-time formalism one can show that
the finite-temperature
\cite{Ma} \cite{BDJ} action for this problem becomes
\bea
S_{TL}& =& \int^{\beta}_{0}d\tau \int dx~ \{p_0 \gamma_0 (\partial_{\tau} -
v_p p_F) \Psi + v_F \bp \gamma_1 \partial_x \Psi\}\nonumber\\
& + &
\int^{\beta}_{0} d\tau \int dx \int dy
\sum_{a,b} \Psi^{\dagger}_a \Psi_a(x,\tau) V_{ab}(x,y) \Psi^{\dagger}_b
\Psi_b(y,\tau).
\label{51}
\eea

For simplicity, we shall set $v_F=1$ and consider the case $v_1 = v_2$
in (\ref{50}) \cite{ML}. We shall also restrict ourselves to the zero
temperature limit ($\beta \rightarrow \infty$). Under these conditions
it is easy to verify that $S_{TL}$ coincides with the non-local Thirring
model discussed in the precedent Sections, provided that the following
identities hold:

\bea
g^2& =& 2\nonumber\\
V_{(0)}(x,y)& =& v_1(x,y) = v_2(x,y)= v( x_1 - y_1)
\delta (x_0 - y_0)\nonumber\\
V_{(1)}& =& 0
\label{52}
\eea

\noindent Of course one has also to make the shift
$\bp\gamma_0 \partial_0 \Psi
\rightarrow \bp\gamma_0 (\partial_0 - p_F ) \Psi$ and identify
$x_0 = \tau $, $x_1 = x$.

One then can employ the method described in the preceding sections in
order to study the Tomonaga-Luttinger model. In \cite{NRT} this model
was studied with emphasis in the bosonization approach. It has also been
previously studied, through a different functional approach, by D.K. Lee
and Y. Chen \cite{LC}. These authors, however, avoided the use of the
decoupling technique applied here. Now we want to examine the vacuum
properties of the model and show how to evaluate the GSWF
considering the model as a special case of a non-local
Thirring model.

Let us first focus our attention to the dispersion relations corresponding
to the elementary excitations of the model at hand. This dispersion
relation is, of course, a special case of (\ref{dr}). Using (\ref{52}) we
have
\be
\omega_{-}^{2}(\bp) = \bp^{2} \{ 1 + \frac{2 v(\bp)}{ \pi} \}
\label{57}
\ee

\noindent which is the well-known result for the spectrum of the
charge-density
excitations of the TL model in the Mattis-Lieb version \cite{ML}.

We can now compute the corresponding GSWF by replacing
(\ref{52}) in (\ref{result2}): 
\be
|\psi_0[\rho]|^2=e^{\frac{1}{2}
\int d\bp 
\rho(\bp)
\sqrt{ (\frac{2 }{\pi  }v(\bp)+1)}
\left(\frac{1}{|\bp|}\right)
\rho(-\bp)}
\ee

For example for $V(x)=1/|x|^2$ we can deduce, employing the same
analysis used for the general model, that the long distance behavior of
the wave functional is essentialy of the Thirring type, with a
renormalization of the exponent of the Jastrow form. On the other hand,
if we consider the $3d$ Coulomb potential $V(\bp)=1/|\bp|^2$, then,
$|\psi|^2\propto exp\{-\mbox{cte}|x-x'| \}$, showing again the landmark
of a confining phase. This means, that no asymptotic fermionic states
can exist, and the dispersion relation (\ref{57}) refers to the
``condensed'' bosonic degrees of freedom. Note that this is a {\em
relativistic bosonic mass mode} ($\omega_{-}^{2}(\bp) = \bp^2 +
\frac{2}{ \pi}$). 

An interesting observation is the following. If we change in (\ref{52})
$V_0\leftrightarrow V_1$, we obtain another model with $j-j$ interaction
rather than $\rho-\rho$ interaction. This new model has the same vacuum
properties of the former (the same GSWF), implying that for the
``Coulomb $j-j$ interaction'' we have also a confining behavior.
However, the confined bosonic modes are very different, since their
dispersion relation is now $\omega_0^2=\bp^4/(\bp^2+\frac{2}{\pi})$.

Another special case that we can analyze is the Sutherland's model
\cite{suther}. It is a system of non-relativistic spinless fermions
interacting via a pair interaction potential $V(|x-y|)$ whose hamilonian
is: 
\be H=\int dx~ \frac{1}{2m}|\partial_x\psi(x)|^2+\frac{1}{2} \int
dx\int dy (\psi^{\dagger}\psi)(x)V(|x-y|)(\psi^{\dagger}\psi)(y)
\label{sou} \ee The spectrum of this hamiltonian was calculated exactly
in \cite{suther}. 

It can be shown that at  long distance, the Sutherland's model 
(\ref{sou}) is equivalent to the model described by the action
\bea
 S_0 &=& \int d^2x~[\bar{\Psi} i
\raise.15ex\hbox {$/$}\kern-.57em\hbox{$\partial$} \Psi
-\frac{g^2}{2}(\bar\psi\gamma_\mu \psi)^2] \nonumber \\
&+&\frac{1}{2}
\int dx dy~ (\psi^{\dagger}\psi)(x)u(|x-y|)(\psi^{\dagger}\psi)(y)
\eea

It is easy to realize that this action is a special case of the one
described by (\ref{free}) and (\ref{int}), provided we identify 
\bea
V_0(x-y)&=&\delta(x-y)-\frac{1}{g^2}u(x-y)\delta(x_0-y_0) \\
V_1(x-y)&=&\delta(x-y) \eea Replacing the Fourier transform of this
expressions in (\ref{result2}) we immediately arrive at 

\be
|\psi_0|^2=e^{               
\frac{1}{2}(\frac{g^2}{\pi}+1)\int d\bp ~
\rho(\bp)\sqrt{  1-\frac{\tilde u(\bp)}{\pi(1+g^2/\pi)} }
\left( \frac{1}{|\bp|}\right) \rho(-\bp) }   
\ee

This wave functional was extensively studied in \cite{FMS}.

\section{GSWF in the Grassmann representation}

Another representation for the wave functional is the so called {\em
Grassmann representation} \cite{LF} \cite{F} \cite{FMS}, in which the
vacuum is projected onto fermionic coherent states. This representation
allows to implement the antisymmetry of the wave functional
automatically. However, the final expression for the functional is less
intuitive than the density representation.

In a subspace with a finite number of particles, we can build a fermionic
coherent state by

\be
|\xi_1,\ldots,\xi_n>=e^{ \sum_{j=1}^{n} \xi_j C^{\dagger}(x_j)}|0>
\label{co-state}
\ee
where $\xi_j$ are Grassmann variables and $C^{\dagger}(x_j)$ are the
fermionic creation operators.

The wave function is constructed by projecting this states onto the
vacuum, i.\ e.\ : 
\be \psi(\xi_1,\ldots,\xi_n)=<0|\xi_1,\ldots,\xi_n>=
<0|e^{ \sum_{j=1}^{n} \xi_j C^{\dagger}(x_j)} |0> 
\label{fgs} 
\ee

We can see the relation between equation (\ref{fgs}) and  the 
{\em orbital} wave functions by expanding the exponential, 

\be \psi(\xi_1,\ldots,\xi_n)=\sum_{n=1}^{N}\frac{1}{(n!)^2} \left(
\prod_{i=1}^{n} \xi_i \right) \psi(x_1,\ldots,x_n) 
\ee
where
$\psi(x_1,\ldots,x_n)=<0|C^{\dagger}(x_1),\ldots,C^{\dagger}(x_n)|0>$.
So, the {\em orbital} wave functions are the coefficients of a
polynomial expansion in the Grassmann variables. 

In the case of a dense system, a coherent state is given by
\be |\chi>=e^{i \int dx~~\chi(x)\hat{\bar{\psi}}(x)+
\hat{\psi}(x)\bar{\chi}(x) }|0> 
\ee
where $\chi(x)$ is a Grassmann field, and a wave functional is labeled by
\be
\psi(\chi,\bar{\chi})=<0|e^{i \int dx~~\chi(x)\hat{\bar{\psi}}(x)+
\hat{\psi}(x)\bar{\chi}(x) }|0> \label{gr-wf} 
\ee

Let us now derive the wave functional (\ref{gr-wf}) for the NLT. The
Grassmann representation of the wave functions can be built in a way
that is quite similar to the one discussed in Section 3 for the density
representation (see Appendix).

It can be shown that the probability for the
state $\chi$ to occur in the ground state is given by

\be
|{\cal{\psi}}_{0}[\bar{\chi},\chi]|^{2} = 
\int D\bar{\eta}~D\eta~Z[\bar{\eta},\eta]~ 
exp (-i \int dx~(\bar{\chi}\eta+\nonumber\\
\bar{\eta}\chi))  \\
\label{01}
\ee
where

\be
Z[\bar{\eta},\eta]  = \int D\bar{\Psi}~D\Psi~e^{-S_0 }\nonumber\\
exp [\int d^2x~(\frac{g^2}{2}J_{\mu}K_{\mu} - 
\bar{\eta}\Psi - \bar{\Psi}\eta)]
\nonumber\\
\ee
and we have taken the equal-time limit $\chi(x,t)=\chi(x)\delta(t)$, and
similarly for $\bar{\chi}$. $S_0$, $J_{\mu}$ and $K_{\mu}$ were defined
in Section 2, and $\eta$ and $\bar{\eta}$ are, of course, a couple of
fermionic sources.

One can rewrite $Z[\bar{\eta},\eta]$, by using the procedure depicted in
Section 2, based on the introduction of a set of auxiliary vector
fields. This leads to an expression which is nothing but the
generalization of equation (\ref{6}) for non-vanishing fermionic
sources:

\begin{equation} 
Z[\bar{\eta},\eta] =  \int DA~ e^{-S'[A]}~exp{[-\int d^2x\bar{\Psi} 
( i\raise.15ex\hbox{$/$}\kern-.57em\hbox{$\partial$} +
\not \!\! A ) \Psi]}~exp{[-\int dx(\bar{\eta}\Psi+\bar{\Psi}\eta)]}
\label{02}
\end{equation}
 
\noindent where $S'[A]$ is given in (\ref{7}).

Performing now a uniform translation in the fields $\Psi$ and
$\bar{\Psi}$, one gets

\begin{equation} 
Z[\bar{\eta},\eta] =  \int DA~ e^{-S'[A]}~ det  
( i\raise.15ex\hbox{$/$}\kern-.57em\hbox{$\partial$} +
\not \!\! A )~ 
exp{-[\int dx dy~ \bar{\eta}(x)~{( i\raise.15ex
\hbox{$/$}\kern-.57em\hbox{$\partial$} +
\not \!\! A )}^{-1}(x,y)~\eta(y)]} 
\label{03}
\end{equation}
 
As explained in Section 3, we can make a chiral transformation in the
fermionic measure and express $A_{\mu}$ in terms of two scalar fields
$\Phi$ and $\omega$ (see equations (\ref{chiral}) and (\ref{longi})).
Taking into account the corresponding Jacobian, we obtain

\begin{equation} 
Z[\bar{\eta},\eta] =  \int D\Phi~D\omega~ e^{-S_{eff}[\Phi,\omega]}~    
{\rm exp}[-\int dx dy~ \bar{\eta}(x)~G_{F}[\Phi,\omega]~\eta(y)] 
\label{04}
\end{equation}

where $S_{eff}[\Phi,\omega]$ picks up the contribution of the Jacobian,
and coincides with (\ref{Sb}) if one sets $Q_{\mu}=0$. We have also
defined

\be
G_{F}[\Phi,\omega] = e^{-g[\gamma_5 \phi(x)+i\eta(x)]}~ 
{( i\raise.15ex\hbox{$/$}\kern-.57em\hbox{$\partial$} )}^{-1}(x,y)~
e^{-g[\gamma_5 \phi(y)-i\eta(y)]}
\label{05}
\ee

At this stage we are ready to insert (\ref{04}) in the expression for
the GSWF, equation (\ref{01}). In so doing one
sees that the integration in the fields $\eta$ and $\bar{\eta}$ is
elementary, yielding

\be
|{\cal{\psi}}_{0}[\bar{\chi},\chi]|^{2} = 
\int D\Phi~D\omega~ e^{-S_{eff}[\Phi,\omega]}~    
{\rm exp}\left[-\int dx dy~ \bar{\chi}(x)~{G_{F}}^{-1}[\Phi,\omega]~
\chi(y)\right]~
\label{06}
\ee
In the functional integrand of (\ref{06}) we have omitted a factor $det~
G_{F}[\Phi,\omega]$, which can be shown to be constant, using, for
instance, a coherent-state definition of the functional integral. 

Now one can expand the exponential and perform the integrations over
$\Phi$ and $\omega$ for each term of the series. The result can be
written as

\begin{equation}
|{\cal{\psi}}_{0}[\bar{\chi},\chi]|^{2} = 
\sum_{n}~\left( \frac{2i}{\pi} \right)^{n}~\frac{1}{n !^{2}}~
\int\left(\prod_{i=1}^{n}dx_{i}dy_{i}\right)\prod_{j=1}^{n}    
{\bar{\chi}}_{\alpha_{j}}(x_{j})~{\chi}_{\beta_{j}}(y_{j})~
F\left([x_j,y_j]\right)_{\alpha_k\beta_k}
\label{07}
\end{equation}
 
\noindent where the indices $\alpha_i$ and $\beta_i$ indicate Dirac
spinor components and $j, k = 1, ..., n$.

The function $F\left([x_j,y_j]\right)_{\alpha_k\beta_k}$ is the product
of two factors:

\begin{equation}
F_0\left([x_j,y_j]\right)_{\alpha_k\beta_k}=(det\frac{1}{(x_i-y_j)})
\prod_{i=1}^n\left(\gamma_1\right)_{\alpha_i\beta_i}~
\label{08}
\end{equation}

\noindent which comes from the contribution of the free fermion
(equal-time) propagators, and a bosonic factor that corresponds to a
multipoint (equal-time) correlation function of vertex operators,

\begin{equation}
{\cal{B}}(x_j,y_j)=\left< {\rm exp}\left[
\sum_{j=1}^n(s_j\Phi(x_j,t)+t_j\Phi(y_j,t))\right]~
{\rm exp}\left[-i\sum_{j=1}^n(\omega(x_j,t)-
\omega(y_j,t))\right]\right>
\label{09}
\end{equation}

\noindent where $s_i~(t_i) = 1$ or $-1$ if $\alpha_i~(\beta_i)=1$ or
$2$. Up to this point the results of this Section are formally equal to
those obtained for the local Thirring model, in ref.\cite{FMS}. However,
we have to stress that in our case the vacuum expectation value in
(\ref{09}) is to be computed for the {\em non-local} model defined by
$S_{eff}[\Phi,\omega]$. Introducing the Fourier transformed fields
$\tilde\Phi(p)$ and $\tilde\omega(p)$, it is straightforward to express
the bosonic factor as
 
\begin{equation}
{\cal{B}}(x_j,y_j)= \int~D\tilde\Phi~D\tilde\omega~
{\rm exp}\left[-\left(S_{eff}
[\tilde\Phi,\tilde\omega]+
\int~d^2p~(\tilde\Phi(p)J(\bp)+\tilde
\omega(p)K(\bp))\right)\right]
\label{010}
\end{equation}

\noindent with
\be
J(\bp)=-\sum_{j=1}^n\left(s_j~e^{i{\bp}x_j}+t_j~e^{i{\bp}y_j}\right)
\label{011}
\ee
\noindent and
\be
K(\bp)=i\sum_{j=1}^n\left(e^{i{\bp}x_j}- e^{i{\bp}y_j}\right)
\label{012}
\ee

As usual, the functional integrations in (\ref{010}) can be easily done
just by conveniently shifting the fields. Indeed, if we introduce two
new fields $\tilde\phi$ and $\tilde\rho$ such that
 
\be
\tilde\Phi(p)=\tilde\phi(p) + M(p) 
\label{013}
\ee
\be
\tilde\omega(p)=\tilde\rho(p) + N(p),
\label{014}
\ee
\noindent the choice

\be
M(-p)=\frac{2B(p)J(\bp)-C(p)K(\bp)}{\Delta(p)} 
\label{015}
\ee
\be
N(-p)=\frac{2A(p)K(\bp)-C(p)J(\bp)}{\Delta(p)}
\label{016}
\ee

\noindent with A,B and C defined in (\ref{seff}) and $\Delta=C^2-4AB$,
allows to obtain

\begin{equation}
{\cal{B}} = {\rm exp}\left[\frac{-1}{(2\pi)^2}\int~d^2p~\frac{1}{\Delta(p)}
\left(B(p) J(\bp)J(-\bp)+A(p) K(\bp)K(-\bp)-C(p)J(\bp)K(-\bp)\right)\right]
\label{017}
\end{equation}

\noindent Please note that in this equation we have omitted the explicit
dependence on the spatial coordinates, which enters the game through $J$
and $K$ (see eqs.(\ref{011}) and (\ref{012})). After replacing the
corresponding expressions the final result can be written in the form

\bea
{\cal{B}}& = &{\rm exp}\left\{(-1/(2\pi)^2)\int~d^2p~(1/\Delta(p))
\sum_{j,k}\left[
(s_js_kB(p)-A(p)+is_jC(p))e^{i{\bp}(x_j-x_k)} \right. \right. \nonumber\\
&+&(t_jt_kB(p)-A(p)-it_jC(p))e^{i{\bp}(y_j-y_k)}+ 
2(s_jt_kB(p)+A(p))\cos[{\bp}(x_j-y_k)]
\nonumber\\
&+&\left. \left. 
iC(p)(t_je^{i{\bp}(y_j-x_k)}-
s_je^{i{\bp}(x_j-y_k)})\right] \right\}
\label{018}
\eea

\noindent This result is to be multiplied by the fermionic factor
$F_0\left([x_j,y_j]\right)_{\alpha_k\beta_k}$ (see eq.(\ref{08})) in
order to find the general term in the series of the squared vacuum
functional, given by eq.(\ref{07}). Thus, we have obtained the general
structure of the ground-state, not only as a functional of the Grassmann
sources ($\bar\chi$,$\chi$) but also of the potentials that bind the
original fermions of the model. Remember that these potentials are
contained in the coefficients A, B and C. If one considers the local
limit, which corresponds to contact interactions in coordinate space,
one can then easily show that eq.(\ref{018}), and therefore also the
general term in (\ref{07}), acquire the Jastrow form, i.e. a factorized
structure with constant exponents given by simple combinations of A,B
and C.

\newpage

\section{Summary and conclusions} In this article we have considered a
recently proposed non-local version of the Thirring model, in which
densities and currents are coupled by arbitrary potential functions
$V_0(|x-y|)$ and $V_1(|x-y|)$, respectively. One of the interesting
aspects of this theory is that it describes, as particular cases,
relevant many-body systems such as the TL and Sutherland's models.

We have focused our attention on the vacuum properties of this model. In
particular, we computed the exact ground-state wave functions, as
functionals of external sources and two-body potentials, in both the
density (Section 3) and Grassmann (Section 5) representations. In the
context of the more intuitive density representation we have stressed
several physical features of the model which, in our framework, can be
easily discussed. For example we got the exact electromagnetic response
of the system for any potential. Concerning the vacuum in itself, we
found a non-trivial symmetry with respect to the interchange of
density-density and current-current potentials. Of course, this symmetry 
does not persist at the level of the dispersion relations of the collective
modes (plasmons), to which the excited states are expected to contribute.
The universal factorized
Jastrow form was also obtained. On the other hand, we have analyzed the
long-distance behavior of the GSWF (Section 4)
for this general model, showing that it contains the TL and Sutherland's
models as special cases. In particular we have examined the asymptotic 
behavior of GSWF's and density-waves frequencies for a wide variety of
power-law potentials. This allowed us to identify different phases contained
in the non-local Thirring model.

We want to emphasize that our results are valid for abitrary bilocal
potentials. This means that in our approach one does not need to specify
the couplings in order to get closed formulae for the GSWF's. Therefore
these formulae could be directly used to obtain the effect of specific
potentials on the vacuum structure. It is also interesting to point out
that the techniques we presented could be easily modified in order to
study the response of the ground-state in the presence of impurities,
following, for example, the lines of ref.\cite{imp}.

\newpage
\appendix
\section{Path integral approach to wave functionals}
\setcounter{equation}{0}
The path integral approach to wave functionals was fully developed in 
refs. \cite{LF}, \cite{F} and \cite{FMS}. 

In order to make this paper self-contained we sketch in this appendix 
the main steps that enable to deduce equations (\ref{gs}) and (\ref{01}). 

\subsection{Density Representation}

In any system with particle number conservation we have 
\be
\partial_t\hat\rho(x,t)+\partial_x\hat j(x,t)=0
\ee
where $\rho$ ($j$) is the charge (current)  density,  
and  a non-relativistic model should  satisfy the following relations:
 \be
[\hat \rho(x), \hat j(x')]=-i\partial_x(\delta(x-x')\hat\rho(x)), 
\ee
\be
[\hat \rho(x),\hat \rho(x') ]=[\hat j(x),\hat j(x') ]=0.
\ee
Thus, we can label the quantum states with the eigenvalues of the 
operator $\hat \rho$, and represent $\hat j$ by 
\be
\hat j(x)|\psi[\rho]>\equiv -i\rho(x)\partial_x\left(
\frac{\delta}{\delta\rho(x)}|\psi[\rho]>
\right)
\ee
This is called the density representation. 

Since this states completely expand all the Hilbert space, we can
resolve the identity operator as 
\be 
\hat I=\int {\cal
D}\rho~|[\rho]><[\rho]| 
\ee

Using this identity, it is easy to derive the relation between wave
functionals and the functional generator of density correlation
functions

\bea
Z(Q_0, Q_1=0) &=&<0|e^{i\int dx~Q_0(x)\hat j_0(x)}|0> \nonumber \\
&=&\int {\cal D}\rho' <0|e^{i\int d^2x~Q_0(x)\hat j_0(x)}|\rho'><\rho'|0> 
\nonumber \\
&=&\int {\cal D}\rho' e^{i\int d^2x~Q_0(x) \rho'(x)}<0|\rho'><\rho'|0>
 \nonumber \\
&=&\int {\cal D}\rho' e^{i\int d^2x~Q_0(x) \rho'(x)}|\psi[\rho']|^2
\eea
Taking the limit for fixed time generators we have
\be
\lim_{Q_0(x)\to Q_0({\bf x})\delta(x_0)} Z(Q_0, Q_1=0) =
\int {\cal D}\rho' e^{i\int d {\bf x} ~Q_0( {\bf x} ) \rho'({\bf x} )}
|\psi[\rho']|^2
\ee
Fourier transforming this expression we finally get  
\bea
\int DQ_0 e^{ -i\int d{\bf x} Q_0({\bf x}) \rho({\bf x}) }
\lim_{Q_0(x)\to Q_0({\bf x})\delta(x_0)}  Z(Q_0, Q_1=0)  &=&   
 \int {\cal D}\rho' \delta(\rho'({\bf x} )-\rho({\bf x})) |\psi[\rho']|^2 
 \nonumber \\
&=& |\psi[\rho]|^2
\eea
which is the equation (\ref{gs}).

\subsection{Grassmann Representation}

Let us begin by considering the generating functional for equal time
fermionic correlation functions

\bea
Z(\eta,\bar{\eta})&=&\int {\cal D}\psi  {\cal D}\bar{\psi} ~
e^{-i S(\bar{\psi},\psi)+i\int dx~\bar{\eta}\psi+\eta\bar{\psi}} 
\nonumber \\
&=&  <0|e^{i\int dx~\bar{\eta}\hat{\psi}+\eta\hat{\bar{\psi}}}|0>
\label{etgf}
\eea
where $\eta(x,x_0)\equiv \eta(x)\delta(x_0)$ is the fermionic source.

The Hilbert space is built through fermionic coherent states:
\be
|\chi,\bar{\chi}>=e^{i \int dx~~\chi(x)\hat{\bar{\psi}}(x)+
\hat{\psi}(x)\bar{\chi}(x) }|0>
\label{cc-ss}
\ee

This space is a complete one, so we can represent the identity operator
as
\be     
I=\int {\cal D}\chi {\cal D}\bar{\chi}~~ |\bar{\chi},\chi><\chi,\bar{\chi}|
\ee
Inserting this expression in (\ref{etgf}) we have
\be
Z(\eta,\bar{\eta})= \int {\cal D}\chi {\cal D}\bar{\chi}
<0|e^{i\int dx~\bar{\eta}\hat{\psi}+\eta\hat{\bar{\psi}}}|\bar{\chi},\chi>
<\chi,\bar{\chi}|0>
\ee

The fermionic operators $\hat{\psi}$ and $\hat{\bar{\psi}}$ satisfy the
following anti-commutation relations
\be
\{\hat\psi_i(x),\hat{\bar{\psi}}_j(y)  \}=(\gamma_0)_{ij}\delta(x-y)
\ee
One can represent these operators by using multiplicative and derivative operators
$\chi_i$ and $\frac{\delta~}{\delta \chi_j}$ acting on the coherent states
(\ref{cc-ss}):  
\bea
\hat{\bar{\psi_i}}&=&\frac{1}{\sqrt{2}}
\left( \bar{\chi}_i+(\gamma_0)_{ij}\frac{\delta~}{\delta\chi_j}
\right)\nonumber \\
\hat{\psi_i}&=&\frac{1}{\sqrt{2}}
\left( \chi_i+(\gamma_0)_{ij}\frac{\delta~}{\delta\bar{\chi}_j}
\right)\nonumber \\
\eea
With this operations we obtain
\be
Z(\eta,\bar{\eta})= \int {\cal D}\chi {\cal D}\bar{\chi}
e^{i\int dx~\bar{\eta}\chi+\eta\bar{\chi}}
|\psi_0(\chi+\gamma_0\eta, \bar{\chi}+\gamma_0\bar{\eta})|^2
\ee
and making the Grassmann translation
\bea
\chi &\longrightarrow& \chi-\gamma_0\eta \nonumber \\
\bar{\chi} &\longrightarrow& \bar{\chi}-\gamma_0\bar{\eta} \nonumber
\eea
we get 
\be
Z(\eta,\bar{\eta})= \int {\cal D}\chi {\cal D}\bar{\chi}
e^{i\int dx~\bar{\eta}\chi+\eta\bar{\chi}}
|\psi_0(\chi, \bar{\chi})|^2
\label{ft}
\ee

Note that formally, (\ref{ft}) is the  
{\em fermionic Fourier Transform} of
$|\psi_0(\chi, \bar{\chi})|^2$. 
We can anti-transform this  expression obtaining finally
\be
|{\cal{\psi}}_{0}[\bar{\chi},\chi]|^{2} = 
\int D\bar{\eta}~D\eta~Z[\bar{\eta},\eta]~ 
exp (-i \int dx~(\bar{\chi}\eta+\nonumber\\
\bar{\eta}\chi)) 
\ee
which is the equation (\ref{01}).

\section*{Acknowledgements} 

This work was partially supported by Conselho Nacional de
Desenvolvimento Cient\'\i fico e Tecnol\'ogico, CNPq (Brasil) and
Consejo Nacional de Investigaciones Cient\'{\i}ficas y T\'ecnicas,
CONICET (Argentina).

\end{document}